\documentclass[10pt,conference]{IEEEtran}
\IEEEoverridecommandlockouts
\usepackage{cite}
\usepackage{amsmath,amssymb,amsfonts}
\usepackage{algorithmic}
\usepackage{graphicx}
\usepackage{textcomp}
\usepackage{xcolor}
\usepackage{subcaption}

\usepackage{amsmath,amssymb,amsfonts}

\usepackage[linesnumbered,ruled,vlined]{algorithm2e}

\usepackage{graphicx}
\usepackage{textcomp}
\usepackage{subcaption}

\usepackage{color,xcolor}
\usepackage{epsfig}
\usepackage{graphicx}
\usepackage{mathtools}

\usepackage{adjustbox}
\usepackage{array}
\usepackage{booktabs}
\usepackage{multirow}

\usepackage{amsmath,amsfonts,amsthm,amssymb}
\usepackage{bm}
\usepackage{nicefrac}
\usepackage{microtype}

\usepackage{changepage}
\usepackage{extramarks}
\usepackage{fancyhdr}
\usepackage{lastpage}
\usepackage{setspace}
\usepackage{soul}
\usepackage{xspace}

\usepackage{marvosym}
\definecolor{BlueViolet}{rgb}{0.26, 0.21, 0.58}
\definecolor{commentcolor}{HTML}{0071bc}
\usepackage[hidelinks,colorlinks,urlcolor=BlueViolet,citecolor=BlueViolet,linkcolor=BlueViolet]{hyperref}
\usepackage[makeroom]{cancel}

\newcommand{\rev}[1]{{#1}}

\newcommand{\cmark}{\checkmark} 

\def\BibTeX{{\rm B\kern-.05em{\sc i\kern-.025em b}\kern-.08em
    T\kern-.1667em\lower.7ex\hbox{E}\kern-.125emX}}
\begin{document}

\title{Distributed Contrastive Learning for \\ Medical Image Segmentation}

\author{Yawen Wu$^1$, Dewen Zeng$^2$, Zhepeng Wang$^3$, Yiyu Shi$^2$, and Jingtong Hu$^1$ \\
$^1$Department of Electrical and Computer Engineering, University of Pittsburgh, USA \\
$^2$Department of Computer Science and Engineering, University of Notre Dame, USA \\
$^3$Department of Electrical and Computer Engineering, George Mason University, USA \\
Email: yawen.wu@pitt.edu, dzeng2@nd.edu, zwang48@gmu.edu, yshi4@nd.edu, jthu@pitt.edu}

\maketitle

\begin{abstract}
Supervised deep learning needs a large amount of labeled data to achieve high performance. However, in medical imaging analysis, each site may only have a limited amount of data and labels, which makes learning ineffective. Federated learning (FL) can learn a shared model from decentralized data. But traditional FL requires fully-labeled data for training, which is very expensive to obtain.
Self-supervised contrastive learning (CL) can learn from unlabeled data for pre-training, followed by fine-tuning with limited annotations. However, when adopting CL in FL, the limited data diversity on each site makes federated contrastive learning (FCL) ineffective.
\rev{In this work, we propose two federated self-supervised learning frameworks for volumetric medical image segmentation with limited annotations. 
The first one features high accuracy and fits high-performance servers with high-speed connections.
The second one features lower communication costs, suitable for mobile devices.}
\rev{In the first framework,} features are exchanged during FCL to provide diverse contrastive data to each site for effective local CL while keeping raw data private. Global structural matching aligns local and remote features for a unified feature space among different sites. 
\rev{In the second framework,}
to reduce the communication cost for feature exchanging, we propose an optimized method FCLOpt that does not rely on negative samples. 
To reduce the communications of model download, we propose the predictive target network update (PTNU) that predicts the parameters of the target network.
Based on PTNU, we propose the distance prediction (DP) to remove most of the uploads of the target network.
Experiments on a cardiac MRI dataset show the proposed \rev{two frameworks} substantially improve the segmentation and generalization performance compared with state-of-the-art techniques.

\end{abstract}

\section{Introduction}

Deep learning (DL) provides state-of-the-art performance for medical applications such as  image segmentation \cite{ronneberger2015u,milletari2016v,xu2019whole,dong2017automatic} and disease diagnosis \cite{liu2020deep, wu2022fairprune} by learning from large-scale labeled datasets, without which the performance of DL will significantly degrade \cite{kairouz2019advances}.
However, medical data exist in isolated medical centers and hospitals \cite{yang2019federated}, and combining a large dataset consisting of very sensitive and private medical data in a single location is impractical and even illegal.
It requires multiple medical institutions to share medical patient data such as medical images, which is constrained by the Health Insurance Portability and Accountability Act (HIPAA) \cite{kairouz2019advances} and EU
General Data Protection Regulation (GDPR) \cite{truong2020privacy}.
Federated learning (FL) is an effective machine learning approach in which distributed clients (i.e. individual medical institutions) collaboratively learn a shared model while keeping private raw data local \cite{rieke2020future,sheller2018multi,sheller2020federated,dou2021federated}.
By applying FL to medical image segmentation, an accurate model can be collaboratively learned and data is kept local for privacy.

\rev{Conventional FL approaches usually use supervised learning on each client} and require that all data are labeled. 
However, annotating all the medical images is usually unrealistic due to the high labeling cost and requirement of expertise.
The deficiency of labels makes supervised FL impractical.
Self-supervised learning can address this challenge by pre-training a neural network encoder with unlabeled data, followed by fine-tuning for a downstream task with limited labels \cite{chaitanya2020contrastive,zeng2021positional}. 
Contrastive learning (CL), a variant of the self-supervised learning approach, can effectively learn high-quality image representations.
By integrating CL to FL as federated contrastive learning (FCL), clients can learn models by first collaboratively learning a shared image-level representation. Then the learned model will be fine-tuned by using limited annotations.
Compared with local CL, FCL can learn a better encoder as the initialization for fine-tuning, and provide higher segmentation performance.
In this way, a high-quality model can be learned by using limited annotations while data privacy is preserved.

\rev{
Based on CL, we propose two frameworks to enable federated self-supervised learning for medical image segmentation. 
The first framework exchanges encoded features for a higher accuracy and has higher communication cost than the second one. It is suitable for distributed medical institutions with high-performance servers and high-speed connections.
The second framework does not exchange encoded features and has less model synchronization. It has lower communication costs and is suitable for mobile devices with high communication costs.
}

Integrating FL with CL to achieve good performance is nontrivial. Simply applying CL to each client and then aggregating the models is not the optimal solution for the following two reasons: 
First, each client only has a small amount of unlabeled data with limited diversity.
Since existing contrastive learning frameworks \cite{chen2020simple,he2020momentum} rely on datasets with diverse data to learn distinctive representations, directly applying CL on each client will result in an inaccurate learned model due to the lack of data diversity.
Second, if each client only focuses on CL on its local data while not considering others' data, each client will have its own feature space based on its raw data and these feature spaces are inconsistent among different clients. 
When aggregating local models, the inconsistent feature space among local models will degrade the performance of the aggregated model.

To address these challenges, \rev{in our \textbf{first} FCL framework, we develop a two-stage FCL method to} enable effective FCL for volumetric medical image segmentation with limited annotations.
The first stage is feature exchange (FE), in which each client exchanges the features (i.e. low-dimensional vectors) of its local data with other clients. It provides more diverse data to compare with for better local contrastive learning while avoiding raw data sharing. In the learning process, the improved data diversity in feature space provides more accurate and complete contrastive information in the local learning process on each client and improves the learned representations. 

The second stage is global structural matching (GSM), in which we leverage structural similarity of 3D medical images to align similar features among clients for better FCL. 
The intuition is that the same anatomical region for different subjects has similar content in volumetric medical images such as MRI.
By leveraging the structural similarity across volumetric medical images, GSM aligns the features of local images to the shared features of the same anatomical region from other clients. In this way, the learned representations of local models are more unified among clients and they further improve the global model after model aggregation.

\rev{
In the first framework, feature exchange requires additional communication. To reduce the communication cost, we further propose the \textbf{second} framework FCLOpt. It is an optimized method that does not rely on negative samples and reduces the communication costs of feature sharing.}
Based on FCLOpt, to further reduce the communications of model download, we propose the predictive target network update (PTNU) that predicts the target network by fast forward.
Based on PTNU, we propose the distance prediction (DP) to remove the upload of the target network.

Experimental results show that the proposed FCL methods substantially improve the segmentation performance over state-of-the-art techniques, and the FCLOpt including the proposed PTNU and DP methods effectively reduces the communication cost while preserving the segmentation performance of FCL.


The rest of this paper is organized as follows. 
The background and related work are described in Section \ref{sect:background}.
The FCL method with feature sharing is introduced in Section \ref{sect:FCL_feature_share}.
The communication-optimized method FCLOpt is described in Section \ref{sect:FCL_reduce_comm}.
The experimental settings and results are reported in Section \ref{sect:exp}, and this paper is concluded in Section \ref{sect:conclusion}.

\section{Background and Related Work}\label{sect:background}
\textbf{Federated Learning.}
Federated learning (FL) learns a shared model by aggregating locally updated models on clients while keeping raw data accessible on local clients for privacy \cite{mcmahan2017communication,li2020federated,zhao2018federated,li2018federated,BaoWXH22}. 
In FL, the training data are distributed among clients.
FL is performed round-by-round by repeating the local model learning and model aggregation process until convergence.
To reduce the communication cost of FL, techniques such as pruning and quantization \cite{jiang2022model,caldas2018expanding,prasad2022reconciling} have been proposed. Sparse learning based methods \cite{bao2020fast,ndiaye2016gap,bao2019efficient,rakotomamonjy2019screening,bao2022accelerated} with theoretical guarantees also have the potential to reduce the communication cost when applied to FL.

The main drawback of these works is that fully labeled data are needed to perform FL, which 
results in high labeling costs.
To solve this problem, 
an FL approach using limited annotations while achieving good performance is needed.

\textbf{Contrastive Learning.} 
Contrastive learning (CL) is a self-supervised approach to learn useful visual representations by using unlabeled data \cite{hadsell2006dimensionality,misra2020self,tian2019contrastive}.
The learned model provides good initialization for fine-tuning on the downstream task with few labels \cite{he2020momentum,chen2020simple,chen2020big,zeng2021positional,chaitanya2020contrastive,wu2021enabling}.
CL performs a proxy task of instance discrimination \cite{wu2018unsupervised}, which maximizes the similarity of representations from similar pairs and minimizes the similarity of representations from dissimilar pairs \cite{wang2020understanding}.

The main drawback of existing CL approaches is that they are designed for centralized learning on large-scale datasets
with sufficient data diversity.
However, when applying CL to FL on each client, the limited data diversity 
will greatly degrade 
the performance of the learned model.
Therefore, an approach to increase the local data diversity while avoiding raw data sharing for privacy is needed.
Besides, while \cite{chaitanya2020contrastive} leverages structural information in medical images for improving centralized CL, it requires accessing raw images of similar pairs for learning.
Since sharing raw medical images is prohibitive due to privacy,
\cite{chaitanya2020contrastive} cannot be applied to FL.
Therefore, an approach to effectively leverage similar images across clients without sharing raw images is needed.

\textbf{Federated Self-supervised Pre-training.}
Some concurrent works employ federated pre-training on unlabeled data.
\rev{
\cite{van2020towards} employs auto-encoders in FL for pre-training on time-series data, but the more effective contrastive learning for visual tasks is not explored in FL.
\cite{bercea2021feddis} uses auto-encoders for federated self-supervised medical image segmentation. 
Different from these works, we employ self-supervised contrastive learning, which has demonstrated superior performance to auto-encoders in centralized training \cite{wu2018unsupervised}.
FedCA \cite{zhang2020federated} combines contrastive learning with FL.
However, it relies on a shared dataset available on each client, which is impractical for medical images due to privacy concerns. Different from this, we do not share raw data among clients to preserve privacy. 
Several works use CL method MoCo to perform self-supervised learning on each client.
In \cite{wu2021federated, wu2022decentralized, wu2021federated_skin}, encoded data representations are shared among clients to improve local CL, while metadata is shared in  \cite{dong2021federated}.
\cite{zhuang2021collaborative,zhuang2022divergence} updates local models of clients adaptively using the exponential moving average (EMA) of the global model.
The proposed work differs from these works in the following ways. First, in our FCL method, we leverage the structural similarity of volumetric images across clients to improve the quality of representation learning. Second, these works only communicate one network between the server and clients even if they use two networks for local learning. Different from this, in our FCLOpt method, we predict the parameters of the second network to achieve higher performance while keeping the communication cost similar to communicating only one network.
}

\section{Federated Contrastive Learning (FCL)}\label{sect:FCL_feature_share}

\subsection{Overview of Federated Contrastive Learning}

\begin{figure*}[!htb]
	\centering
	\includegraphics[width=0.9\textwidth]{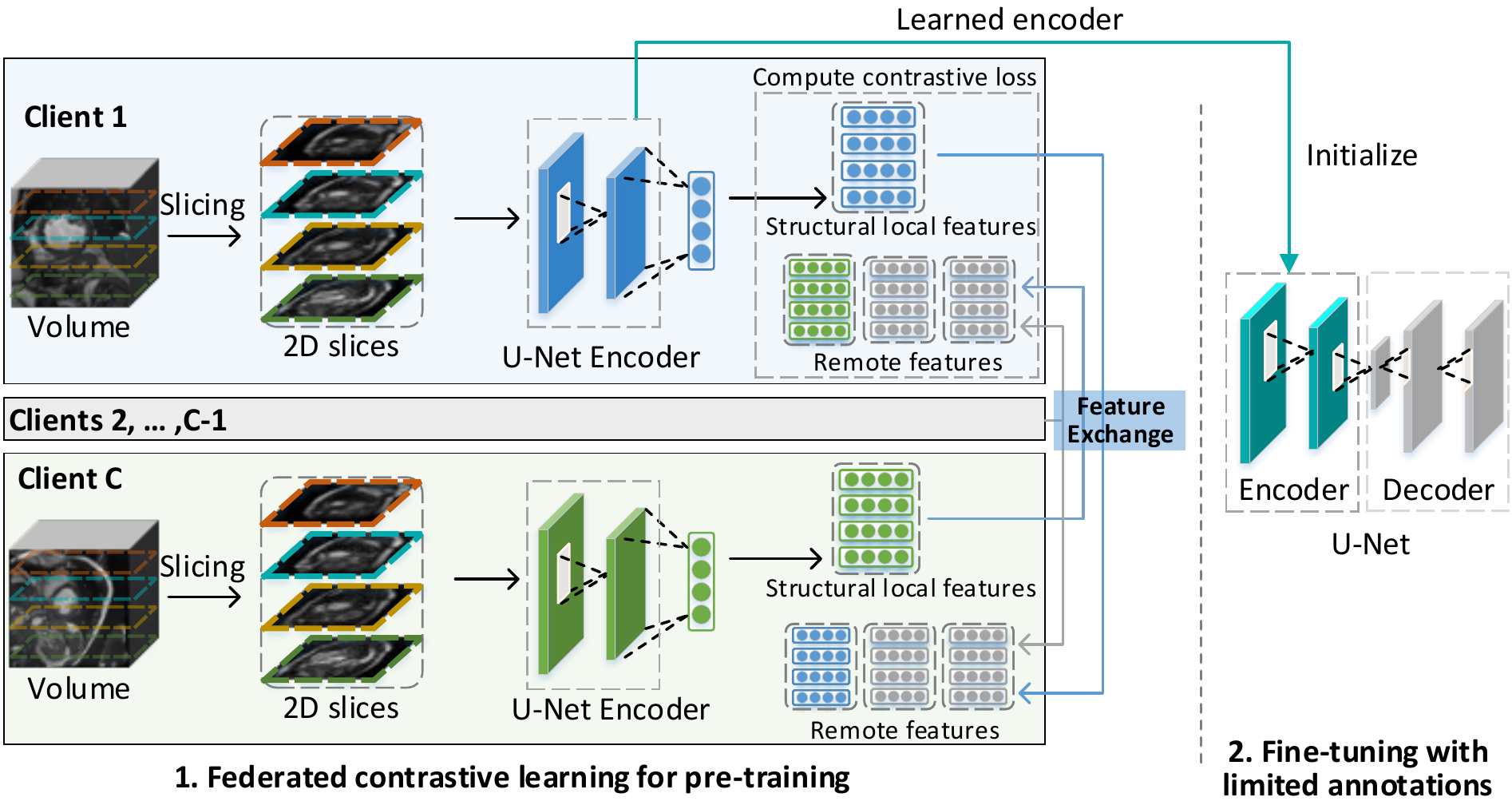}
	\caption{Federated contrastive learning with structural feature exchange for learning the encoder with unlabeled data. 
		Then the learned encoder initializes the encoder in U-Net for fine-tuning with limited annotations.
	}
	\label{fig:overview}
\end{figure*}

The overview of the proposed FCL process is shown in Fig. \ref{fig:overview}. 
Distributed clients first collaboratively learn a shared encoder by FCL with unlabeled data.
Then the learned encoder initializes the encoder in U-Net \cite{ronneberger2015u} for fine-tuning with limited annotations,
either independently on each client by supervised learning or collaboratively by supervised federated learning. 
Since the supervised fine-tuning can be trivially achieved by using available annotations, in the rest of the paper, we focus on FCL to learn a good encoder as the initialization for fine-tuning.

As shown in Fig. \ref{fig:overview}, in the FCL stage, given a volumetric 3D image on one client, multiple 2D slices are sampled from the volume while keeping structural order along the slicing axis. Then the ordered 2D images are fed into the 2D encoder to generate feature vectors, one vector for each 2D image.

To improve the data diversity in local contrastive learning, one natural way is to share raw images \cite{zhao2018federated}. However, sharing raw medical images is prohibitive due to privacy concerns. 
To solve this problem, the proposed FCL framework exchanges the feature vectors instead of raw images among clients, which can improve the data diversity while preserving privacy.
As shown in Fig. \ref{fig:overview}, client $1$ generates structural local features denoted as blue vectors and shares them with other clients. Meanwhile, client $1$ collects structural features from other clients, 
such as remote features shown in green and gray vectors.
After that, the contrastive loss is computed based on both local and remote features.

\subsection{Contrastive Learning with Feature Exchange}

\begin{figure*}[!htb]
	\centering
	\includegraphics[width=1.0\textwidth]{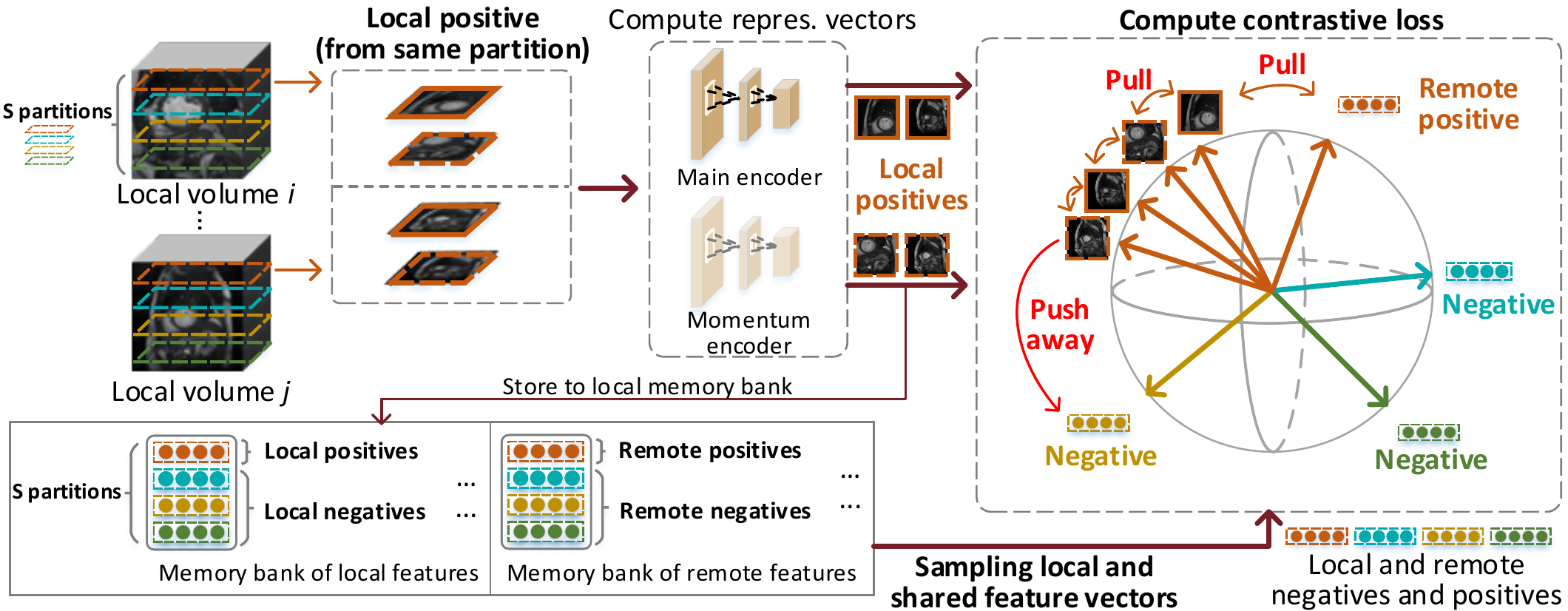}
	\caption{Contrastive learning on one client with exchanged features. The exchanged features consist of remote negatives and remote positives, in which remote negatives improve the local data diversity and remote positives are used for global structural matching to learn a unified feature space among clients.}
	\label{fig:feature_exchange}
\end{figure*}

With feature exchange, each client has both remote and local features and is ready to perform local CL in each round of FCL. The exchanged features from other clients provide more diverse features to compare with and improve the learned representations.
As shown in Fig.~\ref{fig:feature_exchange}, we use MoCo \cite{he2020momentum} architecture for local CL since it has a memory bank for negatives, which can leverage local and remote features. 
There are two encoders, including the main encoder and the momentum encoder.
The main encoder will be learned and used as the initialization for fine-tuning, while the momentum encoder is the slowly-evolving version of the main encoder and generates features to contrast with and for sharing. Now the most important steps are to construct negatives and positives from both local and remote features.

\textbf{Negatives from local and remote features.}
Local features are generated by the momentum encoder from local images and used as local negatives.
Each client has a memory bank of local features and a memory bank of remote features.
Let $Q_{l,c}$ be the size-$K$ memory bank of local features on client $c$, which are used as local negatives.
$Q_{l,c}$ is progressively updated by replacing the oldest features with the latest ones.
In each round of FCL, the remote negatives from other clients will be shared with client $c$ to form its aggregated memory bank including local and remote negatives as:
\begin{equation}\label{equ:shared_negatives}
	Q = Q_{l,c} \cup \{Q_{l,i}\ |\ 1 \le i \le |C|, i\ne c \}.
\end{equation}
where $C$ is the set of all clients and $Q_{l,i}$ is the local memory bank on client $i$.

Compared with using only local memory bank $Q_{l,c}$, the aggregated memory bank $Q$ provides more data diversity to improve CL. 
However, $Q$ is $|C|$ times the size of the local memory bank $Q_{l,c}$. 
More negatives make CL
more challenging since for one local feature $q$, more negatives need to be simultaneously pushed away from it than when using $Q_{l,c}$, which can result in ineffective learning. 
To solve this problem, instead of using all negatives in $Q$, for each $q$ we sample a size-$K$ (i.e. the same size as $Q_{l,c}$) subset of $Q$ as negatives, which is defined as: 
\begin{equation}\label{equ:contrastive_negatives}
	Q^{\prime}=\{ Q_{i} |\ i \sim \mathcal{U}(|Q|,K) \}.
\end{equation}
where $i \sim \mathcal{U}(|Q|,K)$ means $i$ is a set of indices sampled uniformly from $[|Q|]$.

\textbf{Local positives.}
We leverage the structural similarity in the volumetric medical images to define the local positives, in which the same anatomical region from different subjects has similar content \cite{chaitanya2020contrastive}.
Each volume is grouped into $S$ partitions, and one image sampled from partition $s$ of volume $i$ is denoted as $x_s^i$.
\textit{Local positives} are features of images from the same partition in different volumes.
Given an image $x_s^i$, its feature $q_s^i$ and corresponding positives $P(q_s^i)=\{{k_s^i}^{+}, {k_s^j}^{+}\}$ are formed as follows.
Two transformations (e.g. cropping) are applied to $x_s^i$ to get $\tilde{x}_s^i$ and $\hat{x}_s^i$, which are then fed into 
the main encoder and momentum encoder to generate two representation vectors $q_s^i$ and ${k_s^i}^{+}$, respectively.
Then another image $x_s^j$ is sampled from partition $s$ of volume $j$, and its features $q_s^j$ and ${k_s^j}^{+}$ are generated accordingly.
In this way, the local positives for both $q_s^i$ and $q_s^j$ are formed as $P(q_s^i)=P(q_s^j)=\{{k_s^i}^{+}, {k_s^j}^{+}\}$.

\textbf{Loss function for local positives.}
By using the sampled memory bank $Q^{\prime}$ consisting of both \textit{local} negatives and \textit{remote} negatives,
one local feature $q$ is compared with its local positives $P(q)$ and each negative in $Q^{\prime}$. 
The contrastive loss is defined as:
\begin{equation}\label{equ:loss_memorybank}
\begin{split}
	\mathcal{L}_{local} & = \ell_{q,P(q),Q^{\prime}} \\ & = - \frac{1}{|P(q)|} \sum_{k^{+} \in P(q)}  \log \frac{\exp (q \cdot k^{+} / \tau)}{\exp (q \cdot k^{+} / \tau) + \sum_{n\in Q^{\prime}} \exp (q \cdot n / \tau)}.
\end{split}
\end{equation}
where $\tau$ is the temperature and the operator $\cdot$ is the dot product between two vectors.
By minimizing the loss, the distance between $q$ and each local positive is minimized, and the distance between $q$ and each negative in $Q^{\prime}$ is maximized.

\subsection{Global Structural Matching}
\textbf{Remote positives.} We use the remote positives from the shared features to further improve the learned representations.
Inspired by \cite{chaitanya2020contrastive} that aligns the features of images in the same partition for centralized learning, 
on each client, we align the features of one image to the features of images in the same partition from other clients. 
In this way, the features of images in the same partition across clients will be aligned in the feature space and more unified representations can be learned among clients.
To achieve this, for one local feature $q$, in addition to its local positives $P(q)$, we define remote positives $\Lambda(q)$ as features in the sampled memory bank $Q^{\prime}$ which are in the same partition as $q$.
\begin{equation}\label{equ:}
	\Lambda(q) = \{ p \ |\ p \in Q^{\prime}, partition(p)=partition(q) \}.
\end{equation}
$partition(\cdot)$ is the partition number of one feature
and $Q^{\prime}$ is defined in Eq.(\ref{equ:contrastive_negatives}).

\textbf{Final loss function.}
By replacing local positives $P(q)$ in Eq.(\ref{equ:loss_memorybank}) with remote positives $\Lambda(q)$  as $\mathcal{L}_{remote}$,
the final loss function for one feature $q$ is defined as:
\begin{equation}\label{equ:gsm}
	\mathcal{L}_{q} = \mathcal{L}_{remote} + \mathcal{L}_{local} = \ell_{q,\Lambda(q),Q^{\prime}} + \ell_{q,P(q),Q^{\prime}}.
\end{equation}
With $\mathcal{L}_{q}$, the loss for one batch of images is defined as $\mathcal{L}_{B}=\frac{1}{|B|}\sum_{q\in B} \mathcal{L}_{q}$, where $B$ is the set of features generated by the encoder from the batch of images.

\rev{\subsection{Privacy of Feature Exchange}
To protect the shared features against attacks, image encryption methods or representation perturbation methods can be employed.
For the image encryption method, \cite{huang2020instahide} encrypt images before learning and keep the utility of encrypted images for learning.
Images are encrypted before being fed into the encoder for generating representations. As a result, only features of encrypted images are shared, which effectively mitigates the potential vulnerability by feature exchange and maintains the utility of exchanged features for local learning.
For the representation perturbation method, \cite{sun2021soteria} learns to perturb data representation such that the quality of the potentially leaked information is severely degraded, while FL performance is maintained.
}

\begin{figure*}[ht!]
	\centering
	\begin{subfigure}[b]{0.49\linewidth}
		\includegraphics[width=\linewidth]{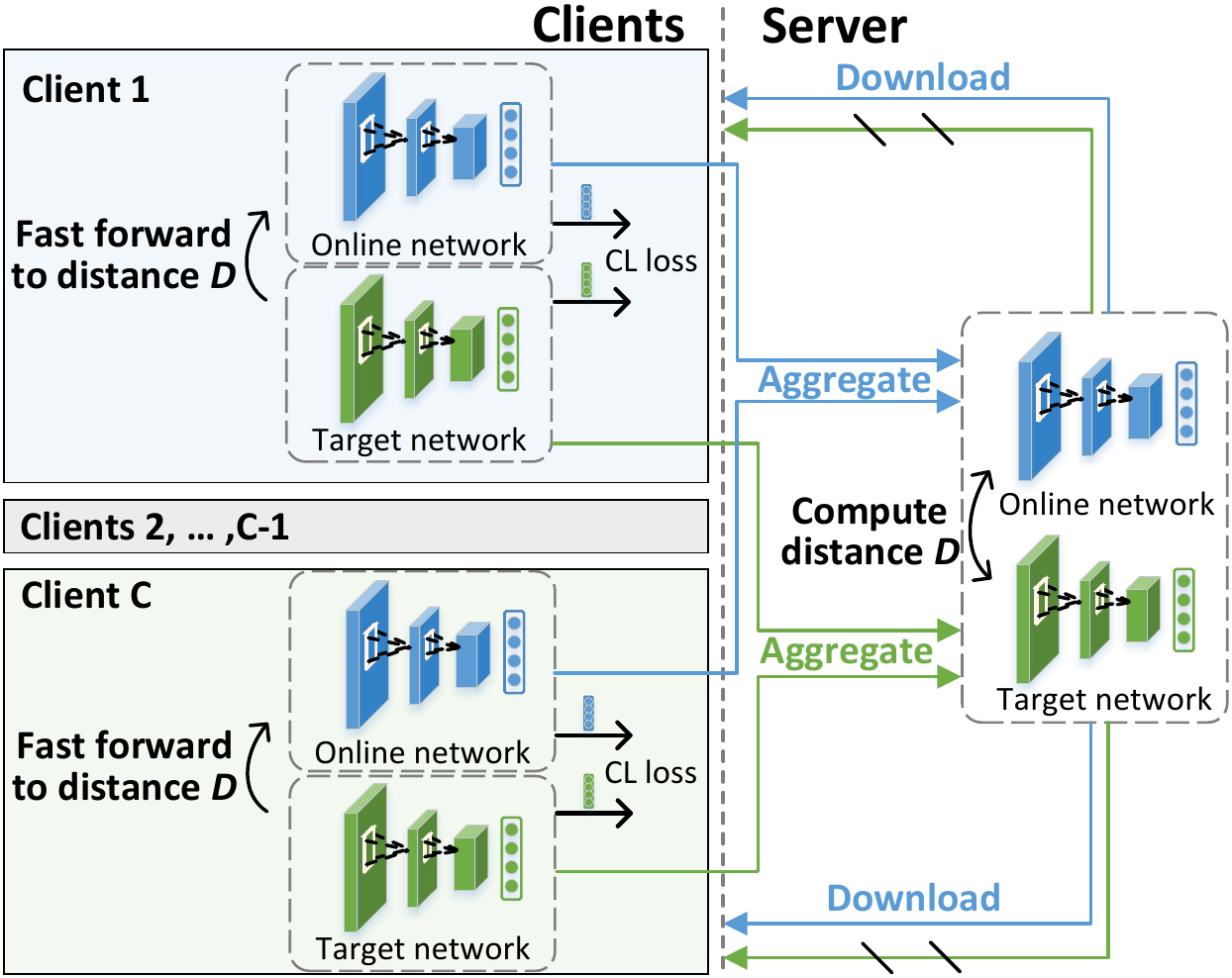}
		\caption{
		FCLOpt with predictive target network update (PTNU).
		The distance $D$ between the global online network and global target network is computed on the server. Each client uses $D$ to predict the global target network instead of downloading it.}
		\label{fig:}
	\end{subfigure}
	\hspace{2pt}
	\begin{subfigure}[b]{0.49\linewidth}
		\includegraphics[width=\linewidth]{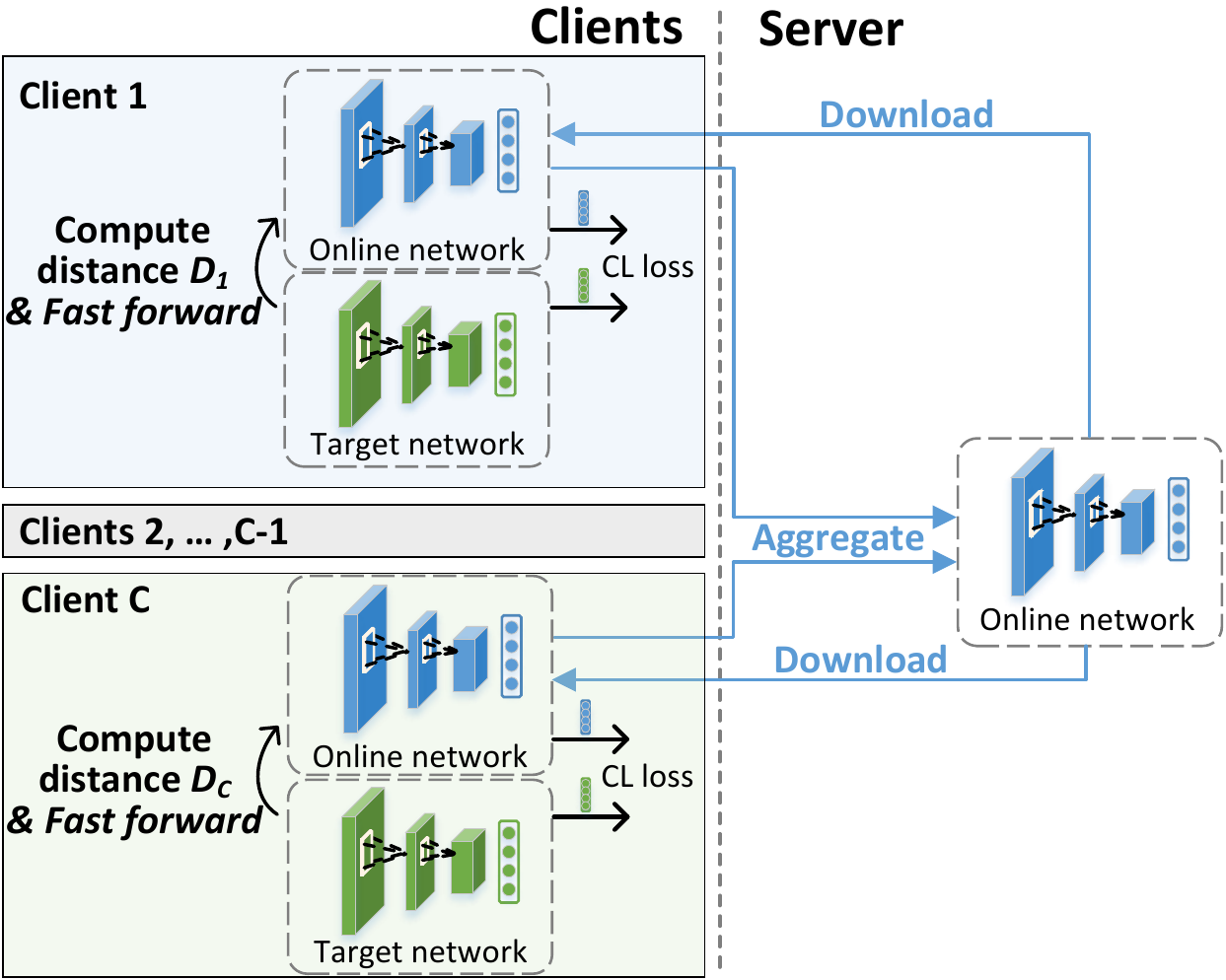}
		\caption{
        FCLOpt-PTNU with distance prediction (DP) between the global online network and the global target network to remove the upload of the target network.}
		\label{fig:}
	\end{subfigure}
	\caption{FCLOpt for reducing communications. 
	Based on the proposed FCL method, we develop FCLOpt to eliminate the need for feature sharing while keeping a high accuracy of the learned model.
	In FCLOpt, we reduce the download and upload of the target encoder. 
	(a) Predictive target network update (PTNU) to eliminate the download of the global target network.
	(b) Distance prediction (DP) to remove most of the uploads of the target network. 
	}
	\label{fig:BYOL_optimize}
\end{figure*}

\section{FCLOpt For Reducing Communications}\label{sect:FCL_reduce_comm}

In the previous Sections, we have introduced the FCL method, aiming at improving the quality of learned representations and the segmentation performance of the downstream task.
However, sharing features require additional communication.
To solve this problem, we eliminate the need for feature sharing by proposing an optimized method \emph{FCLOpt} that does not rely on shared features as negative samples. 
To further reduce the communications of model downloading, we propose the predictive target network update (PTNU) that predicts the target network by fast forward.
Based on PTNU, we propose the distance prediction (DP) to remove most of the uploads of the target network and only use sporadic upload for calibration.

\rev{
\subsection{Revisiting Self-supervised Learning Method BYOL}\label{sect:revisit_byol}
BYOL \cite{grill2020bootstrap} is a self-supervised learning method without negative pairs. 
Conventional CL performs learning by attracting the positive sample pairs and repulsing the negative sample pairs.
Different from this, BYOL directly predicts the output of one sample in a positive pair from the other one.
Since no positive pairs are used in BYOL, it has the potential to eliminate the need for feature exchange and reduces communication cost.}

\rev{
BYOL has a Siamese network architecture, consisting of the online network and the target network.
The online network consists of an encoder and a predictor.
The target network has the same architecture as the encoder in the online network, but different parameters. 
The target network provides the learning targets to train the online network, and it is updated by an exponential moving average (EMA) of the parameters of the online network.
Details of using BYOL for local learning on clients will be introduced in Section \ref{sect:FCLOpt_training}.
}

\subsection{FCLOpt Overview}\label{sect:FCLOpt_overview}
The overview of the optimized method FCLOpt is shown in Fig. \ref{fig:BYOL_optimize}. 
Compared with the FCL method introduced in Section \ref{sect:FCL_feature_share} that solely seeks for high model accuracy, FCLOpt reduces the communication cost, simplifies the system complexity, while keeping a comparable or even better segmentation performance. 

In FCLOpt, there is a server that coordinates multiple clients to upload locally updated online networks and target networks, which are then aggregated on the server as the global online network and the global target network. The server also downloads the global online network and target network to clients.
In each training round, synchronizing both the online network and target network needs extra communication, while synchronizing only one network will greatly degrade the learned model.
To solve this problem, predictive target network update (PTNU) and distance prediction (DP) are proposed to reduce the synchronization of the target network. In this way, the communication cost is comparable to that of synchronizing only one network while the accuracy of the learned network is similar to that of synchronizing two networks.
FedOpt is summarized in Algorithm \ref{algo:fedopt}.

\SetKwInput{KwInput}{Input}                
\SetKwInput{KwOutput}{Output}              
\SetKwFunction{FnClientDistance}{}
\SetKwFunction{FnFastForward}{}
\SetKwFunction{FnClientTrain}{}
\SetKwFunction{FnServer}{}

\begin{algorithm}[ht!]
    \caption{Communication-optimized Federated Contrastive Learning (FCLOpt) with Predictive Target Network Update (PTNU) and Distance Prediction (DP)}
    \label{algo:fedopt}
    \SetAlgoLined
    \KwInput{number of training rounds $R$, number of local epochs $E$, learning rate $\eta$, local batch size $B$, distance calibrator $\alpha$}
    \KwOutput{$F_\theta$}

    \SetKwProg{Fn}{Server}{:}{}
    \Fn{\FnServer{}}{
        Initialize the global online network $F_\theta^0$ and global target network $F_\xi^0$\;
        \For{each round r from 1 to R}{
            $C_r \leftarrow$ (random set of $K$ clients)\;
            \textcolor{commentcolor}{// DP: Predict distance between global online and target networks by Eq.(\ref{equ:dist_pred_raw}) and Eq.(\ref{equ:dist_pred_scaled})}\;
            \For{client $c \in C_r$ in parallel}{
                $dp_c$ = \textbf{ClientDistance}($F_\theta^r$, $r$)\;
            }
            $\tilde{d}$ = $ \alpha \cdot \sum_{c \in C_r} \frac{1}{\lvert C_r \rvert} dp_c$\;
            \For{client $c \in C_r$ in parallel}{
                \textcolor{commentcolor}{//DP eliminates the upload of $f_\xi^c$}\;
                $f_\theta^c, \textcolor{gray}{f_\xi^c} \leftarrow $ \textbf{ClientTrain}($F_\theta^r$, $\tilde{d}$, $r$)\;
            }
            \textcolor{commentcolor}{// Model aggregation of online networks}\;
            $F_\theta^{r+1} \leftarrow \sum_{c\in C_r} \frac{n_c}{n} f_{\theta_c}^r$\;
            \textcolor{commentcolor}{// Model aggregation of target networks. DP eliminates the following two lines}\;
            \textcolor{gray}{$F_\xi^{r+1} \leftarrow \sum_{c\in C_r} \frac{n_c}{n} f_{\xi_c}^r$\;}
            \textcolor{gray}{$\tilde{d} = d(F_\theta^{r+1}, F_\xi^{r+1})$\;}
        }
        \KwRet $f_\phi^{R}$\;
    }
    \SetKwProg{Fn}{ClientDistance}{:}{}
    \Fn{\FnClientDistance{$F_\theta^r$, $r$}}{
        $dist \leftarrow {\lVert F_\theta^{r} - f_{\xi_c}^{r-1} \rVert}_1$\;
        \KwRet $dist$\;
    }
    \SetKwProg{Fn}{ClientTrain}{:}{}
    \Fn{\FnClientTrain{$F_\theta^r$, $\tilde{d}$, $r$}}{
        $f_\theta \leftarrow F_\theta^r$  \textcolor{commentcolor}{// Model download}\;
        \textcolor{commentcolor}{// PTNU: Predict global target network by Algo. \ref{algo:fastforward}}\;
        $F_\xi^r \leftarrow $ \rev{\textbf{PTNU}($F_\theta^r$, $f_\xi^{r-1}$, $\tilde{d}$)}\;
        $\mathcal{B} \leftarrow$ (form batches of size $B$)\;
        \textcolor{commentcolor}{// Local training with Eq.(\ref{equ:byol_cl_loss}) and Eq.(\ref{equ:byol_ema})}\;
        \For{each local epoch i from 1 to E}{
            \For{batch $b \in \mathcal{B}$}{
                $\theta \leftarrow \theta - \eta \triangledown_\theta \mathcal{L_{\theta, \xi}}(\theta; b)$\;
                $\xi \leftarrow \tau \xi + (1-\tau) \theta$\;
            }
        }
        \KwRet $f_\theta$, $\textcolor{gray}{f_\xi}$ \textcolor{commentcolor}{// Model upload}\;
    }
\end{algorithm}

\begin{algorithm}[ht!]
    \caption{Predictive target network update (PTNU).}
    \label{algo:fastforward}
    \SetAlgoLined
    \KwInput{Global online network $F_\theta$, local target network $f_{\xi_c}$, and their distance $d(F_\theta, F_\xi)$}
    \KwOutput{Predicted target model parameters $f_\xi$}
    \SetKwProg{Fn}{PTNU}{:}{}
    \Fn{
        \FnFastForward{$F_\theta$, $f_\xi$, $d$}}{
        \textcolor{commentcolor}{// Compute the initial model distance by Eq.(\ref{equ:distance_exact})}\;
        $dist \leftarrow d (F_\theta, f_{\xi} )$\;
        \While{dist $>$ $d$}{
            \textcolor{commentcolor}{// Update model by exponential moving average (EMA) using Eq.(\ref{equ:approximate_global_target})}\;
            $\xi = m_d \xi + (1-m_d) \theta$\;
            \textcolor{commentcolor}{// Compute updated model distance by Eq.(\ref{equ:distance_exact})}\;
            $dist \leftarrow d (F_\theta, f_{\xi} )$\;
        }
        \KwRet $f_\xi$\;
    }
\end{algorithm}

\subsection{FCLOpt: Local Training and Aggregation}\label{sect:FCLOpt_training}

In FCLOpt, each client has two networks, the online network and the target network, following the model architectures from BYOL \cite{grill2020bootstrap}. 
The online network consists of an online encoder $f_\theta$ and a predictor $q_\theta$. The target network is a target encoder $f_\xi$, which has the same model architecture as the online encoder $f_\theta$ but different parameters.

During local training, two augmentations are applied to a 2D image $x$ and generate two transformed images $t$ and $t^\prime$. 
$t$ and $t^\prime$ are then fed into the online network and the target network, respectively. 
The contrastive loss to update parameters $\theta$ of the online network is defined as:
\begin{equation}\label{equ:byol_cl_loss}
	\mathcal{L}_{\theta, \xi} = {\lVert z-z^\prime \rVert}_2^2 = 2 - 2 \cdot \frac{\langle z, z^\prime \rangle}{{\lVert z \rVert}_2 \cdot {\lVert z^\prime \rVert}_2}.
\end{equation}
where $z=q_\theta (f_\theta (t))$ is the output of the online network, and $z^\prime = f_\xi (t^\prime)$ is the output of the target network. 
No negatives samples are used in this contrastive loss and therefore no shared features are needed.
The online network with parameters $\theta$ is updated by gradient descent to minimize $\mathcal{L}_{\theta, \xi}$, and the target network with parameters $\xi$ is updated by exponential moving average (EMA) of the parameters $\theta$ of the online encoder $f_\theta$:
\begin{equation}\label{equ:byol_ema}
	\xi = m \xi + (1-m) \theta .
\end{equation}
where $m \in (0,1]$ is the momentum parameter controlling the update speed of the target network.
For conciseness, in the rest of this paper, we use $f_\theta$ to denote the whole online network consisting of the online encoder and the online predictor.

On each client, the training is performed for $E$ epochs before uploading the updated local networks to the server for aggregation and downloading the aggregated models from the server to initiate the networks for training in the next round.

The aggregation is performed on the server as follows.
In round $r$, denoting the online network and target network after local training on client $c$ as $f_{\theta_c}^r$ and $f_{\xi_c}^r$, 
the global online network $F_\theta^{r+1}$ and the global target network $F_\xi^{r+1}$ 
are aggregated as:
\begin{equation}\label{equ:byol_online_agg}
	F_\theta^{r+1} = \sum_{c\in C} \frac{n_c}{n} f_{\theta_c}^r .
\end{equation}
\begin{equation}\label{equ:byol_target_agg}
	F_\xi^{r+1} = \sum_{c\in C} \frac{n_c}{n} f_{\xi_c}^r .
\end{equation}
where $n_c$ is the number of samples on client $c$, and $n$ is the total number of samples on all clients.
After network aggregation, $F_\theta^{r+1}$ and $F_\xi^{r+1}$ are downloaded to clients to start the training of round $r+1$.

\subsection{Predictive Target Network Update}\label{sect:PTNU}

Since there are two networks to synchronize between the server and clients, to reduce the communication cost of the target network, we propose a predictive target network update (PTNU) method to eliminate the need for target network download.

We first introduce how to eliminate the download of the global target network by predicting the parameters $\xi$ of global target network $F_\xi^{r+1}$ on clients.

At the beginning of round $r+1$, the server computes the average $\ell_1$ distance between the parameters of the aggregated $F_\theta^{r+1}$ and $F_\xi^{r+1}$:
\begin{equation}\label{equ:distance_exact}
	d(F_\theta^{r+1}, F_\xi^{r+1}) = {\lVert F_\theta^{r+1} - F_\xi^{r+1} \rVert}_1 = \frac{1}{N} \sum_{a \in \theta,\ b \in \xi} {\lvert a-b \rvert}.
\end{equation}
where $N = {\lVert \xi \rVert}_0$ is the number of parameters in the global target network.

On client $c$, ideally, both $F_\theta^{r+1}$ and $F_\xi^{r+1}$ are downloaded from the server to initiate its local $f_{\theta_c}^{r+1}$ and $f_{\xi_c}^{r+1}$. 
To reduce the communication cost, we only download $F_\theta^{r+1}$ and the scalar value $d(F_\theta^{r+1}, F_\xi^{r+1})$, and predict $F_\xi^{r+1}$ on the client instead of downloading it.
\rev{Following Algorithm \ref{algo:fastforward},}
given the latest global online network $F_\theta^{r+1}$, the local target network $f_{\xi_c}^{r}$, and the distance $d(F_\theta^{r+1}, F_\xi^{r+1})$, we predict the parameters of $F_\xi^{r+1}$ by an iterative update on client $c$:
\begin{equation}\label{equ:approximate_global_target}
	\xi_c = m_d \xi_c + (1-m_d) \theta^{r+1}
\end{equation}
which is performed iteratively until $d(F_\theta^{r+1}, f_{\xi_c}^{r}) \le d(F_\theta^{r+1}, F_\xi^{r+1})$ is satisfied. 
After this update, the parameters of $f_{\xi_c}^{r}$ are used to initialize $F_\xi^{r+1}$ on client $c$ to start the training of round $r+1$.

The intuition of PTNU is that we can approximate $F_\xi^{r+1}$ by gradually approaching $F_\theta^{r+1}$ from a certain direction until a certain frontier is reached.
Ideally, $F_\xi^{r+1}$ is the weighted average of all $f_{\xi_c}^r, c \in C$ as defined in Eq.(\ref{equ:byol_online_agg}), 
and $F_\theta^{r+1}$ is the weighted average of all $f_{\theta_c}^r, c \in C$ as defined in Eq.(\ref{equ:byol_target_agg}).
For each $c \in C$, since $f_{\xi_c}^r$ is updated by EMA in Eq.(\ref{equ:byol_ema}) to approach $f_{\theta_c}^r$ from the direction of $f_{\xi_c}^r$, $F_\xi^{r+1}$ can be treated as approaching $F_\theta^{r+1}$ from the direction of each $f_{\xi_c}^r$ simultaneously.

On client $c$, since we only know $f_{\xi_c}^r$ and do not have access to $f_{\xi_i}^r, i \ne c$, we gradually approach $F_\theta^{r+1}$ from the direction of $f_{\xi_c}^r$, instead of from the direction of each $f_{\xi_c}^r$.
By updating $f_{\xi_c}^{r}$ with Eq.(\ref{equ:approximate_global_target}), we draw $f_{\xi_c}^{r}$ near to $F_\theta^{r+1}$ until their distance is comparable to the distance between $F_\theta^{r+1}$ and $F_\xi^{r+1}$.
In this way, the updated $f_{\xi_c}^{r}$ can approximate $F_\xi^{r+1}$.

The PTNU eliminates the need for downloading the aggregated global target network $F_\xi^{r+1}$ to clients. Considering the communications of four components, including upload of the local online network $f_{\theta_c}^r$ and the local target network $f_{\xi_c}^r$, and the download of the global online network $F_\theta^{r+1}$ and the global target network $F_\xi^{r+1}$, PTNU reduces about 25\% of the communication.

\subsection{Distance Prediction for Predictive Target Network Update}\label{sect:DP}

To further reduce the communication cost, we propose a distance prediction (DP) method that eliminates most of the uploads of the local target network $f_{\xi_c}^r$ and only uses sporadic upload for calibration.
In this way, by combining DP with PTNU, most of the communications regarding the target networks are removed, which can reduce about 50\% of the communications.

To perform PTNU by Eq.(\ref{equ:approximate_global_target}), the exact distance $d(F_\theta^{r+1}, F_\xi^{r+1})$ between the aggregated online network $F_\theta^{r+1}$ and the target network $F_\xi^{r+1}$ is required, which is computed on the server by Eq.(\ref{equ:distance_exact}).
Since $F_\xi^{r+1}$ is aggregated by the uploaded local target models $f_{\xi_c}^r, c \in C$ with Eq.(\ref{equ:byol_target_agg}), to reduce the upload of $f_{\xi_c}^r$, instead of computing the exact $d(F_\theta^{r+1}, F_\xi^{r+1})$ on the server, we approximate it by a proxy distance $\tilde{d}(F_\theta^{r+1}, F_\xi^{r+1})$ computed on the clients.

At the beginning of round $r+1$, each client $c \in C$ downloads $F_\theta^{r+1}$ from the server, and 
computes $dp_c$ as:
\begin{equation}
    dp_c = d(F_\theta^{r+1}, f_{\xi_c}^r) = {\lVert F_\theta^{r+1} - f_{\xi_c}^r \rVert}_1
\end{equation}
which is then uploaded to the server to predict the distance $\tilde{d}(F_\theta^{r+1}, F_\xi^{r+1})$ as:

\begin{equation}\label{equ:dist_pred_raw}
    DP = \frac{1}{\lvert C \rvert} \sum_{c \in C } dp_c.
\end{equation}
\begin{equation}\label{equ:dist_pred_scaled}
	\tilde{d}(F_\theta^{r+1}, F_\xi^{r+1}) = \alpha DP.
\end{equation}
Compared with the exact distance $d(F_\theta^{r+1}, F_\xi^{r+1})=d(F_\theta^{r+1}, \sum_{c\in C} \frac{n_c}{n} f_{\xi_c}^r)$, which computes the weighted average of local target networks $f_{\xi_c}^r$ on the server before computing the $\ell_1$ distance, 
the distance prediction first computes the distance between the global online network $F_\theta^{r+1}$ and the target network $f_{\xi_c}^r$ on the client, after which the distance averaging is taken on the server.
Since the value of $DP$ and the ground truth distance $d(F_\theta^{r+1}, F_\xi^{r+1})$ are slightly different, 
we use a calibration parameter $\alpha$, which is slightly smaller than 1, to accommodate for the difference between $DP$ and $d(F_\theta^{r+1}, F_\xi^{r+1})$.
In this way, $\tilde{d}(F_\theta^{r+1}, F_\xi^{r+1})$ as the adjusted $DP$ can accurately approximate the $d(F_\theta^{r+1}, F_\xi^{r+1})$.

$dp_c$ is computed on each client directly from local target networks $f_{\xi_c}^r$ without the need for uploading $f_{\xi_c}^r$.
After computing $d(F_\theta^{r+1},f_{\xi_c}^r)$ on each client $c$, this distance is uploaded to the server for the averaging in Eq.(\ref{equ:dist_pred_raw}) and scaling in Eq.(\ref{equ:dist_pred_scaled}).
Then the value of $\tilde{d}(F_\theta^{r+1}, F_\xi^{r+1})$ is downloaded to clients and used as the target distance defined in Eq.(\ref{equ:distance_exact}) for performing PTNU.

In the training process, the scaling factor between $d(F_\theta^{r+1}, F_\xi^{r+1})$ and $DP$ can gradually shift and the calibration parameter $\alpha$ needs to be adjusted to make the prediction $\tilde{d}(F_\theta^{r+1}, F_\xi^{r+1})$ closely follow the real distance $d(F_\theta^{r+1}, F_\xi^{r+1})$. 
To achieve this, in every $R$ rounds, we periodically upload the local target networks to the server for computing the ground-truth distance $d(F_\theta^{r+1}, F_\xi^{r+1})$ by Eq.(\ref{equ:distance_exact}), and adjust the calibration parameter as:
\begin{equation}\label{equ:}
	\alpha = \frac{d(F_\theta^{r+1}, F_\xi^{r+1})} {DP} .
\end{equation}
where $DP$ is the distance prediction by Eq.(\ref{equ:dist_pred_raw}). 
By using the calibrated $\alpha$, we can perform accurate distance prediction by Eq.(\ref{equ:dist_pred_scaled}) for the following training rounds, which further helps the PTNU process.

\section{Experiments}\label{sect:exp}

\textbf{Dataset and preprocessing.}
We evaluate the proposed approaches on the ACDC MICCAI 2017 challenge dataset \cite{bernard2018deep}.
It consists of 100 patients with 3D cardiac MRI images.
Each patient has about 15 volumes covering a full cardiac cycle, and only volumes for the end-diastolic and end-systolic phases are annotated by experts for three structures, including the left ventricle, myocardium, and right ventricle.
The HVSMR MICCAI 2016 challenge dataset \cite{pace2015interactive} contains 10 3D cardiac MRI images captured in an axial view using a 1.5T scanner with expert annotations of the blood pool and ventricular myocardium.
In the pre-processing, for both datasets, following \cite{chaitanya2020contrastive}
we first normalize the intensity of each 3D volume $x$ using min-max normalization to $[x_1,x_{99}]$, where $x_p$ is the $p^{th}$ intensity percentile in $x$. 
Then we resample the 2D images and corresponding annotations to a fixed pixel size $r_f=1.25 \times 1.25mm^2$
and $r_f=0.7 \times 0.7mm^2$ for ACDC and HVSMR, respectively.

\noindent
\textbf{Federated and training setting.}
Following \cite{zhao2018federated}, we use 10 clients. 
We randomly split 100 patients in ACDC dataset into 10 partitions, each with 10 patients.
Then each client is assigned one partition with 10 patients.
We use the proposed FCL approaches to pre-train the U-Net encoder on the assigned dataset partition on each client without labels.
Then the pre-trained encoder (i.e. the final global encoder after pre-training) is used as the initialization for fine-tuning the U-Net segmentation model by using a small number of labeled samples.
The U-Net model follows the standard 2D U-Net architecture \cite{ronneberger2015u} with the initial number of channels set to 48.
We evaluate with three settings for fine-tuning: \textit{local fine-tuning}, \textit{federated fine-tuning}, and \textit{centralized fine-tuning}.
In local fine-tuning, each client fine-tunes the model on its local annotated data.
In federated fine-tuning, all clients collaboratively fine-tune the model by supervised FL with a small number of annotations.
In centralized fine-tuning, all data are combined for sampling the annotated data, following a standard evaluation protocol \cite{caron2020unsupervised, chen2020simple}.

\noindent
\textbf{Evaluation.}
During fine-tuning, we use 5-fold cross-validation to evaluate the segmentation performance. In each fold, 10 patients on one client are split into a training set of 8 patients and a validation set of 2 patients. 
For each fold, we fine-tune with annotations from $N\in \{1,2,4,8\}$ patients in the training set, and validate on the validation set of the same fold on all clients (i.e. 20 patients). 
Dice similarity coefficient (DSC) is used as the metric for evaluation. 

\noindent
\textbf{Training details.}
The FCL is performed for 200 rounds. 
The percentage of active clients per round is 1.0 and the number of local epochs per communication round is 1.
The size of the memory bank is 4096. The temperature $\tau$ for contrastive loss is 0.1 and the momentum is 0.99.
The SGD optimizer is used with momentum 0.9 and weight decay 0.0001.
The batch size is 32 and the learning rate is 0.05 with a cosine decay schedule.
For FCLOpt, FCLOpt-PTNU, and FCLOpt-PTNU-DP, the learning rate is 0.5 with a cosine decay schedule.
The momentum parameter $m$ for the target network update is 0.99.
$m_d$ for PTNU is 0.995. 
The calibration parameter $\alpha$ is adjusted every 10 training rounds.
As in \cite{chaitanya2020contrastive}, we group each volume into 4 partitions.
In the fine-tuning stage, the model is trained for 200 epochs in local fine-tuning or 200 rounds in federated fine-tuning.
Adam optimizer is used with a batch size of 10, a learning rate of 0.0005, and a cosine schedule.
The training is performed on one Nvidia V100 GPU.

\noindent
\textbf{Baselines.}
We compare the proposed approaches with multiple baselines.
\textit{Random init} fine-tunes the model from random initialization.
\textit{Local CL} performs contrastive learning on each client by the SOTA approach \cite{chaitanya2020contrastive} with unlabeled data for pre-training the encoder before fine-tuning.
\textit{Rotation} \cite{gidaris2018unsupervised} is a self-supervised pre-training approach by predicting the image rotations.
\textit{SimCLR} \cite{chen2020simple}, \textit{SwAV} \cite{caron2020unsupervised}, \textit{MoCo} \cite{he2020momentum}, and \textit{BYOL} \cite{grill2020bootstrap} are the SOTA \rev{self-supervised learning} approaches for pre-training.
We combine these three self-supervised approaches with \textit{FedAvg} \cite{mcmahan2017communication} as their federated variants \textit{FedRotation}, \textit{FedSimCLR}, \textit{FedSwAV}, \textit{FedMoCo}, and \textit{FedBYOL} for pre-training the encoder.
\rev{
\textit{FedGL} is the combination of the SOTA self-supervised learning approach for volumetric medical image segmentation \cite{chaitanya2020contrastive} with \textit{FedAvg}.
\textit{FedCA} \cite{zhang2020federated} and \textit{FedU} \cite{zhuang2021collaborative} are two federated self-supervised learning methods for pre-training.}
In the experimental results, we denote the method introduced in Section \ref{sect:FCL_feature_share} as \textit{FCL}, the method described in Section \ref{sect:FCL_reduce_comm} without PTNU or DP as \textit{FCLOpt}, and denote the methods with PTNU and DP enabled as \textit{FCLOpt-PTNU} and \textit{FCLOpt-PTNU-DP}, respectively.

\subsection{Results of Local Fine-tuning}\label{sect:exp_local_finetune}

\begin{table}[!htb]
	\centering
	\caption{Comparison of the proposed approaches and baselines on \textbf{local fine-tuning} with limited annotations on the ACDC dataset. $N$ is the number of annotated patients for fine-tuning on each client. 
		The average dice score and standard deviation across 10 clients are reported, in which on each client the dice score is averaged on 5-fold cross-validation. 
		The proposed approaches substantially outperform all the baselines with different numbers of annotations. 
	}
	\label{tab:exp_local_finetune}
	\setlength\tabcolsep{3.0pt}
	\resizebox{1.0\columnwidth}{!}{
		\begin{tabular}{lcccc}
			\toprule
			Methods     & $N$=1    & $N$=2    & $N$=4    & $N$=8    \\ \midrule
			Random init & 0.280 $\pm$ 0.037 & 0.414 $\pm$ 0.070 & 0.618 $\pm$ 0.026 & 0.766 $\pm$ 0.027 \\
			Local CL \cite{chaitanya2020contrastive}   & 0.320 $\pm$ 0.106 & 0.456 $\pm$ 0.095 & 0.637 $\pm$ 0.043 & 0.770 $\pm$ 0.029 \\
			FedRotation \cite{gidaris2018unsupervised} & 0.357 $\pm$ 0.058 & 0.508 $\pm$ 0.054 & 0.660 $\pm$ 0.021 & 0.783 $\pm$ 0.029\\
			FedSimCLR \cite{chen2020simple}  & 0.288 $\pm$ 0.049 & 0.435 $\pm$ 0.046 & 0.619 $\pm$ 0.032 & 0.765 $\pm$ 0.033 \\
			FedSwAV \cite{caron2020unsupervised}    & 0.323 $\pm$ 0.066 & 0.480 $\pm$ 0.067 & 0.659 $\pm$ 0.019 & 0.782 $\pm$ 0.030 \\
			FedCA \cite{zhang2020federated}      & 0.280 $\pm$ 0.047 & 0.417 $\pm$ 0.042 & 0.610 $\pm$ 0.030 & 0.766 $\pm$ 0.029 \\
			FedMoCo \cite{he2020momentum} & 0.287 $\pm$ 0.056 & 0.442 $\pm$ 0.066 & 0.626 $\pm$ 0.034 & 0.767 $\pm$ 0.030  \\
			FedBYOL \cite{grill2020bootstrap} & 0.431 $\pm$ 0.057 & 0.554 $\pm$ 0.052 & 0.685 $\pm$ 0.021 & 0.781 $\pm$ 0.027   \\
			\rev{FedU \cite{zhuang2021collaborative}} &   \rev{\underline{0.441} $\pm$ 0.047} & \rev{\underline{0.586} $\pm$ 0.043} & \rev{\underline{0.703} $\pm$ 0.018} & \rev{\underline{0.795} $\pm$ 0.022}    \\
			\rev{FedGL \cite{chaitanya2020contrastive}}  &  \rev{0.260 $\pm$ 0.036} & \rev{0.404 $\pm$ 0.063} & \rev{0.633 $\pm$ 0.028} & \rev{0.765 $\pm$ 0.040}   \\ 
			FCL (ours)    & 0.506 $\pm$ 0.056 & 0.631 $\pm$ 0.051 & \textbf{0.745} $\pm$ 0.017 & \textbf{0.824} $\pm$ 0.025 \\
			FCLOpt (ours)    & \textbf{0.524} $\pm$ 0.052 & \textbf{0.655} $\pm$ 0.039 & \textbf{0.745} $\pm$ 0.020 & 0.821 $\pm$ 0.020 \\
			FCL-PTNU (ours) & 0.517 $\pm$ 0.061 & 0.622 $\pm$ 0.045 & 0.730 $\pm$ 0.019 & 0.810 $\pm$ 0.022 \\
			FCL-PTNU-DP (ours) & 0.512 $\pm$ 0.053 & 0.621 $\pm$ 0.050 & 0.729 $\pm$ 0.016 & 0.810 $\pm$ 0.027 \\
			\bottomrule
		\end{tabular}
	}
\end{table}

We evaluate the performance of the proposed approaches by fine-tuning locally on each client with limited annotations.
As shown in Table \ref{tab:exp_local_finetune}, the proposed approaches substantially outperform the baselines.
\rev{First, with 1, 2, 4, or 8 annotated patients, our FCL method
outperforms the best-performing baseline by 0.065, 0.045, 0.042, and 0.029 dice score, respectively.}
Our communication-optimized methods FCLOpt, FCL-PTNU, and FCL-PTNU-DP achieve a similar or higher dice score than our method FCL. 
Second, the proposed approaches significantly improve the annotation efficiency.
For example, 
\rev{with 2 or 4 annotated patients, our FCLOpt method performs on par with the best-performing baseline with 2$\times$ annotations (0.655 vs. 0.703 and 0.745 vs. 0.795), respectively.}

\subsection{Results of Federated Fine-tuning}\label{sect:exp_federated_finetune}

\begin{table}[!htb]
	\centering
	\caption{Comparison of the proposed approaches and baselines on \textbf{federated fine-tuning} with limited annotations on the ACDC dataset. \rev{$N$ is the number of annotated patients for fine-tuning on each client. $L$ is the total number of annotated patients from all clients. }
		The proposed approaches significantly outperform all the baselines with different numbers of annotations.}
	\label{tab:exp_federated_finetune}
	\setlength\tabcolsep{3.0pt}
	\resizebox{1.0\columnwidth}{!}{
		\begin{tabular}{lcccc}
			\toprule
			\rev{Annotated patients per client}     & $N$=1    & $N$=2    & $N$=4    & $N$=8    \\ 
			\rev{Annotated patients of all clients}     & \rev{L=$1\times10$}    & \rev{L=$2\times10$}    & \rev{L=$4\times10$}    & \rev{L=$8\times10$}    \\ \midrule
			Random init & 0.445 $\pm$ 0.012 & 0.572 $\pm$ 0.061 & 0.764 $\pm$ 0.017 & 0.834 $\pm$ 0.011 \\
			Local CL \cite{chaitanya2020contrastive}   & 0.473 $\pm$ 0.013 & 0.717 $\pm$ 0.024 & 0.784 $\pm$ 0.015 & 0.847 $\pm$ 0.009 \\
			FedRotation \cite{gidaris2018unsupervised} & 0.516 $\pm$ 0.015 & 0.627 $\pm$ 0.074 & 0.821 $\pm$ 0.015 & 0.867 $\pm$ 0.010 \\
			FedSimCLR \cite{chen2020simple}  & 0.395 $\pm$ 0.023 & 0.576 $\pm$ 0.046 & 0.788 $\pm$ 0.014 & 0.859 $\pm$ 0.011 \\
			FedSwAV \cite{caron2020unsupervised}    & 0.500 $\pm$ 0.015 & 0.594 $\pm$ 0.058 & 0.815 $\pm$ 0.015 & 0.862 $\pm$ 0.010 \\
			FedCA \cite{zhang2020federated}      & 0.397 $\pm$ 0.020 & 0.561 $\pm$ 0.047 & 0.784 $\pm$ 0.015 & 0.858 $\pm$ 0.011 \\
		    FedMoCo \cite{he2020momentum} & 0.467 $\pm$ 0.016 & 0.675 $\pm$ 0.053 & 0.782 $\pm$ 0.018 & 0.846 $\pm$ 0.011 \\
		    FedBYOL \cite{grill2020bootstrap} & \underline{0.621} $\pm$ 0.065 & \underline{0.790} $\pm$ 0.011 & 0.840 $\pm$ 0.006 & 0.871 $\pm$ 0.006  \\
			\rev{FedU \cite{zhuang2021collaborative}} &  \rev{0.576 $\pm$ 0.082} & \rev{0.717 $\pm$ 0.105} & \rev{\underline{0.850} $\pm$ 0.010} & \rev{\underline{0.879} $\pm$ 0.007}   \\
			\rev{FedGL \cite{chaitanya2020contrastive}} &  \rev{0.468 $\pm$ 0.019} & \rev{0.687 $\pm$ 0.051} & \rev{0.813 $\pm$ 0.015} & \rev{0.865 $\pm$ 0.009   }\\ 
			FCL (ours)   & 0.646 $\pm$ 0.052 & 0.824 $\pm$ 0.004 & 0.871 $\pm$ 0.007 & 0.894 $\pm$ 0.006 \\
			FCLOpt (ours)   & \textbf{0.769} $\pm$ 0.025 & \textbf{0.853} $\pm$ 0.006 & \textbf{0.877} $\pm$ 0.006 & \textbf{0.897} $\pm$ 0.004 \\ 
			FCL-PTNU (ours) & 0.680 $\pm$ 0.086 & 0.840 $\pm$ 0.007 & 0.868 $\pm$ 0.009 & 0.889 $\pm$ 0.006 \\
			FCL-PTNU-DP (ours) & 0.778 $\pm$ 0.016 & 0.840 $\pm$ 0.002 & 0.873 $\pm$ 0.006 & 0.894 $\pm$ 0.004 \\
			\bottomrule
		\end{tabular}
	}
\end{table}

We evaluate the performance of the proposed approaches by collaborative federated fine-tuning with limited annotations.
Similar to local fine-tuning, the proposed approaches significantly outperform the SOTA techniques as shown in Table \ref{tab:exp_federated_finetune}. 
First, with 1, 2, 4, or 8 annotated patients per client (i.e. 10, 20, 40, or 80 annotated patients in total), \rev{our FCLOpt method 
outperforms the best-performing baselines by 0.148, 0.063, 0.027, and 0.018 dice score, respectively.}
Second, the proposed approaches effectively reduce the annotations needed for fine-tuning. For example, with 2 or 4 annotated patients per client, \rev{our FCLOpt method achieves better performance than the best-performing baseline with 2$\times$ annotated patients per client (0.853 vs. 0.850 and 0.877 vs. 0.879, respectively)},
which achieve more than 2$\times$ labeling-efficiency.
Third, compared with local fine-tuning in Table \ref{tab:exp_local_finetune}, all the approaches achieve a higher dice score. 
This is because federated fine-tuning with annotations on distributed clients leverages more annotations than local fine-tuning with only local annotations.

\subsection{Results of Centralized Fine-tuning}\label{sect:exp_centralized_finetune}

\begin{table}[!htb]
	\centering
	\caption{Comparison of the proposed approaches and baselines on \textbf{centralized fine-tuning} with limited annotations on the ACDC dataset. $N$ is the number of annotated patients for fine-tuning. 
		The proposed approaches significantly outperform all the baselines with different numbers of annotations.}
	\label{tab:exp_centralized_finetune}
	\setlength\tabcolsep{3.0pt}
	\resizebox{1.0\columnwidth}{!}{
		\begin{tabular}{lcccc}
			\toprule
			Methods     & $N$=1    & $N$=2    & $N$=4    & $N$=8    \\ \midrule
			Random init & 0.296 $\pm$ 0.091 & 0.528 $\pm$ 0.064 & 0.677 $\pm$ 0.056 & 0.797 $\pm$ 0.028 \\
			Local CL \cite{chaitanya2020contrastive} & 0.314 $\pm$ 0.058 & 0.544 $\pm$ 0.065 & 0.691 $\pm$ 0.040 & 0.805 $\pm$ 0.014 \\
			FedRotation \cite{gidaris2018unsupervised} & 0.374 $\pm$ 0.072 & 0.583 $\pm$ 0.061 & 0.686 $\pm$ 0.056 & 0.815 $\pm$ 0.021 \\
			FedSimCLR \cite{chen2020simple}  & 0.287 $\pm$ 0.030 & 0.524 $\pm$ 0.065 & 0.658 $\pm$ 0.037 & 0.802 $\pm$ 0.022 \\
			FedSwAV \cite{caron2020unsupervised}    & 0.334 $\pm$ 0.096 & 0.575 $\pm$ 0.063 & 0.726 $\pm$ 0.044 & 0.805 $\pm$ 0.020 \\
			FedCA \cite{zhang2020federated}      & 0.320 $\pm$ 0.067 & 0.527 $\pm$ 0.067 & 0.653 $\pm$ 0.049 & 0.793 $\pm$ 0.024 \\
			FedMoCo \cite{he2020momentum} & 0.310 $\pm$ 0.068 & 0.520 $\pm$ 0.055 & 0.695 $\pm$ 0.045 & 0.802 $\pm$ 0.017  \\
			FedBYOL \cite{grill2020bootstrap} & 0.472 $\pm$ 0.073 & \underline{0.633} $\pm$ 0.042 & 0.729 $\pm$ 0.039 & 0.805 $\pm$ 0.021  \\
			\rev{FedU \cite{zhuang2021collaborative}} &  \rev{\underline{0.485} $\pm$ 0.133} & \rev{\underline{0.633} $\pm$ 0.057} & \rev{\underline{0.747} $\pm$ 0.047} & \rev{\underline{0.820} $\pm$ 0.031}   \\
			\rev{FedGL \cite{chaitanya2020contrastive}} &  \rev{0.331 $\pm$ 0.057} & \rev{0.520 $\pm$ 0.066} & \rev{0.686 $\pm$ 0.045} & \rev{0.795 $\pm$ 0.020}   \\ 
			FCL (ours)    & 0.575 $\pm$ 0.113 & 0.702 $\pm$ 0.041 & \textbf{0.790} $\pm$ 0.026 & \textbf{0.844} $\pm$ 0.024 \\
			FCLOpt (ours)    & \textbf{0.587} $\pm$ 0.109 & \textbf{0.708} $\pm$ 0.055 & 0.785 $\pm$ 0.038 & 0.837 $\pm$ 0.031 \\
			FCL-PTNU (ours) & 0.547 $\pm$ 0.130 & 0.691 $\pm$ 0.052 & 0.771 $\pm$ 0.041 & 0.837 $\pm$ 0.020 \\
			FCL-PTNU-DP (ours) & 0.556 $\pm$ 0.132 & 0.674 $\pm$ 0.069 & 0.763 $\pm$ 0.055 & 0.831 $\pm$ 0.031 \\
			\bottomrule
		\end{tabular}
	}
\end{table}

We evaluate the performance of the proposed approaches by centralized fine-tuning with limited annotations, which is a standard evaluation protocol for generic self-supervised learning.
As shown in Table \ref{tab:exp_centralized_finetune}, the proposed approaches {FCL} and {FCLOpt} achieve significantly higher performance than the baselines.
First, with 1, 2, 4, or 8 annotated patients, \rev{our {FCL} method outperforms the best-performing baseline by 0.090, 0.069, 0.043, and 0.024 dice score}, respectively, while our communication-optimized {FCLOpt}, FCL-PTNU, and FCL-PTNU-DP perform on par with {FCL}.
Second, all our methods significantly improve the annotation efficiency.
For example, 
with 2 or 4 annotated patients, \rev{{FCL} performs on par with the best-performing baseline with 2$\times$ annotations (0.702 vs. 0.747 and 0.790 vs. 0.820), respectively}, which roughly improves labeling-efficiency by 2$\times$.

\begin{table}[!htb]
	\centering
	\caption{Impact of learning rate on the random init baseline in the centralized fine-tuning setting. $N$ is the number of annotated patients for fine-tuning. 
	}
	\label{tab:random_init_lr}
	\resizebox{1.0\columnwidth}{!}{
		\begin{tabular}{ccccc}
			\toprule
			LR     & $N$=1    & $N$=2    & $N$=4    & $N$=8    \\ \midrule
			\textbf{0.0005 (default)} & 0.296 $\pm$ 0.091 & 0.528 $\pm$ 0.064 & 0.677 $\pm$ 0.056 & 0.797 $\pm$ 0.028 \\
			0.0010 & 0.311 $\pm$ 0.035 & 0.546 $\pm$ 0.062 & 0.686 $\pm$ 0.037 & 0.813 $\pm$ 0.028 \\
			0.0020 & 0.294 $\pm$ 0.035 & 0.519 $\pm$ 0.080 & 0.680 $\pm$ 0.044 & 0.798 $\pm$ 0.027 \\
			0.0050 & 0.280 $\pm$ 0.033 & 0.485 $\pm$ 0.055 & 0.666 $\pm$ 0.069 & 0.795 $\pm$ 0.029 \\
			0.0100 & 0.290 $\pm$ 0.069 & 0.480 $\pm$ 0.043 & 0.668 $\pm$ 0.055 & 0.806 $\pm$ 0.017 \\
			\bottomrule
		\end{tabular}
	}
\end{table}

\rev{
In addition to the default learning rate of 0.0005, we further explore more learning rates for the random init baseline in the centralized fine-tuning setting.
As shown in Table \ref{tab:random_init_lr}, increasing the learning rate only results in marginal improvement and even degrades the performance of the random init baseline when the learning rate is too large. Therefore, the default learning rate we used is a good one. Besides, we use the same learning rate for all the baselines for fine-tuning, which is a fair comparison. 
}

\subsection{Results of Transfer Learning}\label{sect:exp_transfer_learning}

\begin{table}[!htb]
	\centering
	\caption{Comparison of the proposed methods and baselines on \textbf{transfer learning} from ACDC to HVSMR dataset. $M$ is the number of annotated patients for fine-tuning. The average dice score and standard deviation on 5-fold cross-validation are reported. The proposed approaches outperform all the baselines, which shows the proposed approaches can learn useful and transferable representations to be used on the downstream task.}
	\label{tab:exp_transfer_acdc_hvsmr}
	\setlength\tabcolsep{3.0pt}
	\resizebox{1.0\columnwidth}{!}{
    \begin{tabular}{lcccc}
    \toprule
    Methods     & $M$=1    & $M$=2    & $M$=4    & $M$=8    \\ \midrule
    Random init & 0.792 $\pm$ 0.051 & 0.814 $\pm$ 0.049 & 0.827 $\pm$ 0.049 & 0.859 $\pm$ 0.039 \\
    Local CL \cite{chaitanya2020contrastive}  & 0.798 $\pm$ 0.053 & 0.811 $\pm$ 0.045 & 0.825 $\pm$ 0.052 & 0.855 $\pm$ 0.044 \\
    FedRotation \cite{gidaris2018unsupervised}  & 0.800 $\pm$ 0.054 & 0.816 $\pm$ 0.055 & 0.834 $\pm$ 0.052 & 0.864 $\pm$ 0.037 \\
    FedSimCLR \cite{chen2020simple}  & 0.797 $\pm$ 0.048 & 0.799 $\pm$ 0.048 & 0.815 $\pm$ 0.053 & 0.854 $\pm$ 0.040 \\
    FedSwAV \cite{caron2020unsupervised}   & 0.802 $\pm$ 0.044 & 0.814 $\pm$ 0.054 & 0.842 $\pm$ 0.039 & 0.862 $\pm$ 0.040 \\
    FedCA \cite{zhang2020federated}  & 0.790 $\pm$ 0.043 & 0.802 $\pm$ 0.050 & 0.817 $\pm$ 0.056 & 0.861 $\pm$ 0.037 \\
    FedMoCo \cite{he2020momentum} & 0.794 $\pm$ 0.049 & 0.815 $\pm$ 0.043 & 0.828 $\pm$ 0.045 & 0.857 $\pm$ 0.039  \\
    FedBYOL \cite{grill2020bootstrap} & 0.797 $\pm$ 0.047 & 0.802 $\pm$ 0.042 & 0.834 $\pm$ 0.045 & \underline{0.865} $\pm$ 0.031  \\
	\rev{FedU \cite{zhuang2021collaborative}} &  \rev{\underline{0.806} $\pm$ 0.039} & \rev{\underline{0.819} $\pm$ 0.042} & \rev{\underline{0.843} $\pm$ 0.040} & \rev{0.862  $\pm$ 0.047}    \\
	\rev{FedGL \cite{chaitanya2020contrastive}} & \rev{0.791 $\pm$ 0.054} & \rev{0.813 $\pm$ 0.049} & \rev{0.827 $\pm$ 0.054} & \rev{0.860 $\pm$ 0.032}   \\ 
	FCL (ours)    & \textbf{0.814} $\pm$ 0.046 & 0.823 $\pm$ 0.048 & \textbf{0.849} $\pm$ 0.038 & \textbf{0.872} $\pm$ 0.033 \\ 
	FCLOpt (ours)   & \textbf{0.814} $\pm$ 0.045  & \textbf{0.828} $\pm$ 0.039 & 0.844 $\pm$ 0.042 & \textbf{0.872} $\pm$ 0.028 \\
	FCL-PTNU (ours) & 0.812 $\pm$ 0.040 & 0.829 $\pm$ 0.039 & 0.843 $\pm$ 0.044 & 0.868 $\pm$ 0.032  \\
	FCL-PTNU-DP (ours) & 0.813 $\pm$ 0.040 & 0.832 $\pm$ 0.048 & 0.847 $\pm$ 0.040 & 0.868 $\pm$ 0.032  \\
    \bottomrule
    \end{tabular}
    }
\end{table}

We evaluate the generalization performance of the learned encoder by transferring to a new downstream task. We pre-train the encoder on ACDC by different methods and use the pre-trained encoder as the initialization for fine-tuning on the HVSMR dataset with limited annotations.
The results are shown in Table \ref{tab:exp_transfer_acdc_hvsmr}. Under different numbers of annotations for fine-tuning, the proposed approaches consistently outperform the baselines.
While the acquisition view and resolutions are different on the source ACDC dataset and target HVSMR dataset, 
the proposed approaches can still learn useful and transferable representations to be used on the downstream task.

\subsection{Ablation Studies of FCL}

\begin{table}[!htb]
	\centering
	\caption{Ablation study of FCL on the ACDC dataset. The average dice score and standard deviation across 10 clients by local fine-tuning are reported. $N$ is the number of annotated patients for fine-tuning on each client. }
	\label{tab:ablation_fcl}
	\resizebox{1.0\columnwidth}{!}{
    \begin{tabular}{ccccc}
    \toprule
    Methods     & $N$=1    & $N$=2    & $N$=4    & $N$=8    \\ \midrule
    Without FE & 0.287 $\pm$ 0.056 & 0.442 $\pm$ 0.066 & 0.626 $\pm$ 0.034 & 0.767 $\pm$ 0.030 \\
    FE & 0.296 $\pm$ 0.048 & 0.445 $\pm$ 0.069 & 0.634 $\pm$ 0.035 & 0.768 $\pm$ 0.028 \\
    FE+NS & 0.373 $\pm$ 0.057 & 0.524 $\pm$ 0.064 & 0.678 $\pm$ 0.021 & 0.787 $\pm$ 0.027 \\
    FCL (FE+NS+GSM)    & \textbf{0.506} $\pm$ 0.056 & \textbf{0.631} $\pm$ 0.051 & \textbf{0.745} $\pm$ 0.017 & \textbf{0.824} $\pm$ 0.025 \\ \bottomrule
    \end{tabular}
    }
\end{table}

We perform ablation studies to evaluate the effectiveness of each component in the FCL with feature sharing introduced in Section \ref{sect:FCL_feature_share}.
The influences of feature exchange (FE) by Eq.(\ref{equ:shared_negatives}), negative sampling (NS) by Eq.(\ref{equ:contrastive_negatives}), and global structural matching (GSM) by Eq.(\ref{equ:gsm}) on federated contrastive learning are evaluated. 
By progressively adding the proposed FE, NS, and GSM, the average dice score increases, which shows the effectiveness of each of the proposed approaches.
As shown in Talbe \ref{tab:ablation_fcl}, by using a given number of annotations for fine-tuning, enabling the proposed components FE, NS, and GSM one by one effectively improves the dice score after fine-tuning.
For example, with 4 annotated patients, adding FE+NS improves the dice score from 0.626 to 0.678, and adding GSM further improves the dice score to 0.745.
These results show the effectiveness of each component of FCL.

\subsection{Results of Reduced Communication Cost}\label{sect:exp_reduced_comm}

\begin{table}[!htb]
	\centering
	\caption{Comparison of the proposed approaches and baselines on \textbf{local fine-tuning} with limited annotations on the ACDC dataset. \emph{Communication} is the amount of data communications in the FCL pretraining process.
	$N$ is the number of annotated patients for fine-tuning on each client. 
	}
	\label{tab:exp_comm}
	\setlength\tabcolsep{3.0pt}
	\resizebox{1.0\columnwidth}{!}{
		\begin{tabular}{lccccc}
			\toprule
			Methods & Comm. & $N$=1    & $N$=2    & $N$=4    & $N$=8    \\ \midrule
\multicolumn{6}{c}{\textit{Baselines}} \\
			FedMoCo & 0.544 $\times$ & 0.287 $\pm$ 0.056 & 0.442 $\pm$ 0.066 & 0.626 $\pm$ 0.034 & 0.767 $\pm$ 0.030  \\
			FedBYOL & 0.373 $\times$ & 0.431 $\pm$ 0.057 & 0.554 $\pm$ 0.052 & 0.685 $\pm$ 0.021 & 0.781 $\pm$ 0.027 \\
\hline
\multicolumn{6}{c}{\textit{Our methods: high accuracy with feature sharing}} \\
			FCL (ours) &          1.000 $\times$   & \textbf{0.506} $\pm$ 0.056 & \textbf{0.631} $\pm$ 0.051 & \textbf{0.745} $\pm$ 0.017 & \textbf{0.824} $\pm$ 0.025 \\
\hline
\multicolumn{6}{c}{\textit{Our methods: optimizing communication}} \\
			FCLOpt (ours) &     0.645 $\times$ & \textbf{0.524} $\pm$ 0.052 & \textbf{0.655} $\pm$ 0.039 & \textbf{0.745} $\pm$ 0.020 & \textbf{0.821} $\pm$ 0.020 \\
			FCL-PTNU (ours) &   0.509 $\times$ & 0.517 $\pm$ 0.061 & 0.622 $\pm$ 0.045 & 0.730 $\pm$ 0.019 & 0.810 $\pm$ 0.022 \\
			FCL-PTNU-DP (ours) & 0.386 $\times$ & 0.512 $\pm$ 0.053 & 0.621 $\pm$ 0.050 & 0.729 $\pm$ 0.016 & 0.810 $\pm$ 0.027 \\
			\bottomrule
		\end{tabular}
	}
\end{table}

We evaluate the effectiveness of \emph{FCLOpt}, \emph{PTNU}, and \emph{DP} for reducing the communication cost while keeping a high segmentation performance, which are introduced in Section \ref{sect:FCL_reduce_comm}.
The results are shown in Table \ref{tab:exp_comm}, where the results of fine-tuning locally on each client are reported.
We report the amount of data communication in each round of the federated pre-training, and the segmentation performance after fine-tuning.
The communication cost is normalized such that the method FCL has a cost of $1.0$ and standards for 124.2 MB communication per round.
First, all of our four methods FCL (without communication optimization), FCLOpt, FCL-PTNU, and FCL-PTNU-DP achieve a high segmentation performance compared with the baselines FedMoCo and FedBYOL.
We compare with the baselines FedMoCo and FedBYOL because our FCL employs the same base CL method MoCo \cite{he2020momentum} as FedMoCo, and FCLOpt, FCL-PTNU, FCL-PTNU-DP use the same base CL method BYOL \cite{grill2020bootstrap} as FedBYOL.
Second, compared with FCL, our communication-optimized FCLOpt effectively reduces the communication cost from $1.000 \times$ to $0.645 \times$. 
Adding $PTNU$ to FCLOpt as FCL-PTNU further reduces the communication cost to $0.509\times$ 
and adding $DP$ reduces the communication cost to $0.386\times$.
Compared with FedBYOL which only synchronizes the online network, our FCL-PTNU-DP has a comparable communication cost but significantly better segmentation performance.
These results show all of our methods achieve a high segmentation performance, and enabling the communication optimizations can effectively reduce the communication cost.

\begin{figure*}[ht!]
\centering
  \includegraphics[width=0.9\linewidth]{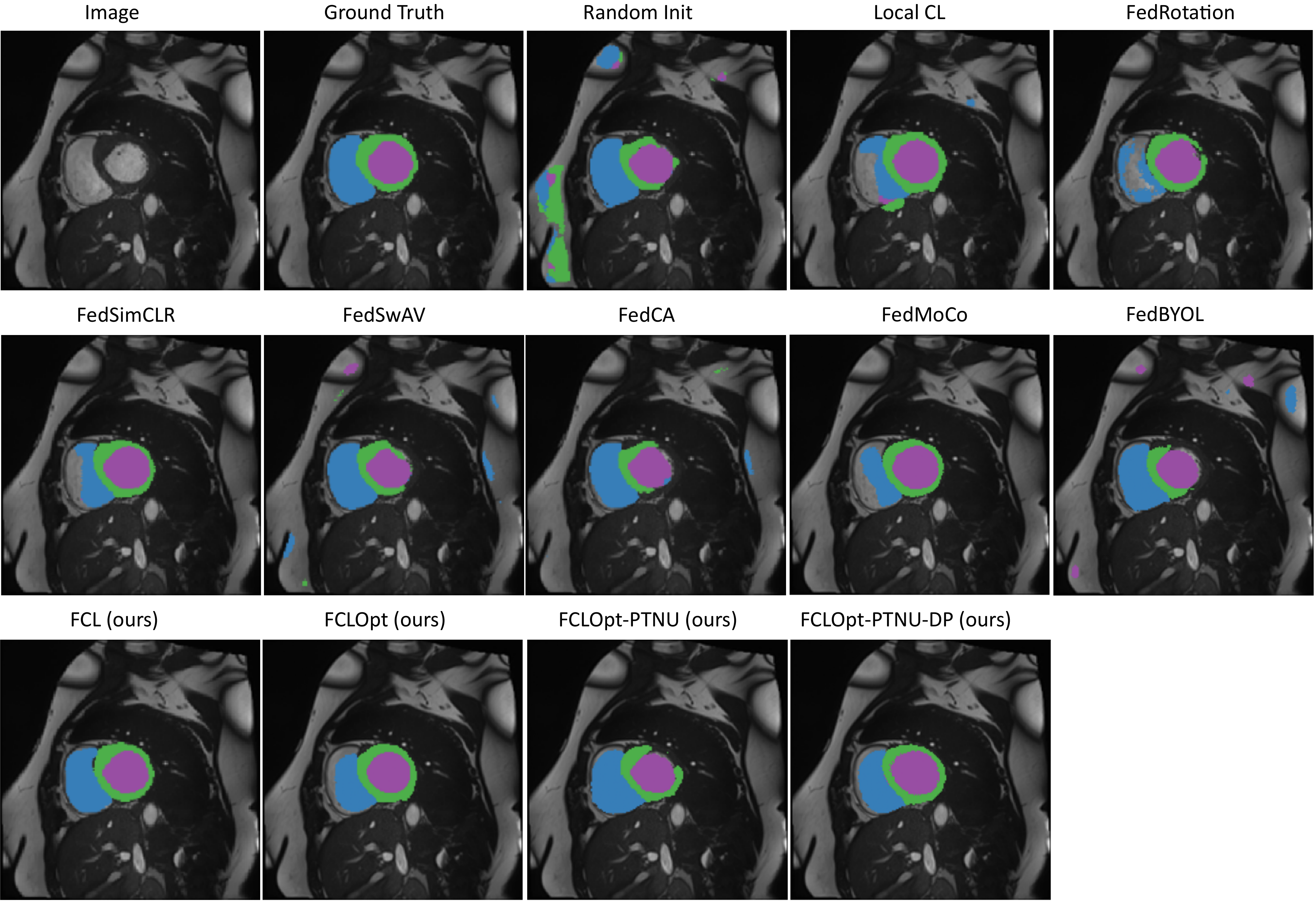}
\caption{Visualization of segmentation results on ACDC dataset. The results are generated from the fine-tuned model when 2 annotated patients are used for fine-tuning ($N=2$). The proposed approaches achieve significantly better segmentation performance than the baselines.}
\label{fig:visualize}
\end{figure*}

\subsection{Comparison of Communicated Model Components}
\begin{table}[!htb]
	\centering
	\caption{Communicated model components in different methods. 
	}
	\label{tab:sync_components}
	\setlength\tabcolsep{3.0pt}
	\resizebox{1.0\columnwidth}{!}{
    \begin{tabular}{lc cc  cc  c } 
    \toprule 
    \multirow{2}{*}{Methods} & Online & Online & Target & Encoded  \\ 
    & network & predictor & network & representations  \\ \midrule
    \multicolumn{5}{c}{\textit{MoCo-based methods}} \\
    FedMoCo & \cmark &  & \cmark &  & \\
    FCL (ours) & \cmark &  & \cmark & \cmark \\ \midrule
    \multicolumn{5}{c}{\textit{BYOL-based methods}} \\
    FedBYOL & \cmark & \cmark &  & \\
    FCLOpt (ours) & \cmark & \cmark & \cmark &  & \\
    FCLOpt-PTNU (ours) & \cmark & \cmark & Upload only &  & \\
    FCLOpt-PTNU-DP (ours) & \cmark & \cmark & $~\sim 0$  &  & \\ \bottomrule
    \end{tabular}
}
\end{table}
To better understand different methods, we show the communicated model components in Table \ref{tab:sync_components}, in which the top half compares MoCo-based methods, while the bottom half compares BYOL-based methods.
First, both MoCo-based baseline FedMoCo and our FCL communicate the online network and the target network (the online predictor does not exist in MoCo). The difference is that our FCL also communicates the encoded features. As shown in Table \ref{tab:exp_comm}, our FCL greatly outperforms FedMoCo in terms of model performance at the cost of increased communication cost. This is desirable when model performance is the main goal while communication has a marginal cost for medical institutions with high-speed connections.
Second, for BYOL-based methods, the goal of our FCLOpt with PTNU and DP is to achieve a similar communication cost as FedBYOL while having substantially higher model performance. More specifically, in FedBYOL, only the online network and the online predictor are communicated, while the target network is not. Based on FedBYOL, we propose FCLOpt which further communicates the target network for higher model performance. Then, we propose PTNU which eliminates the upload of the target network. After that. we propose DP to eliminate most of the downloads of the target network. As shown in Table \ref{tab:exp_comm}, our FCLOpt-PTNU-DP achieves much higher model performance than FedBYOL and has almost the same communication cost as FedBYOL.

\subsection{Visualization}\label{sect:exp_visualization}

We visualize the segmentation results of the ACDC dataset in Fig. \ref{fig:visualize}. The input image and the ground truth annotations are shown in the first two images, followed by segmentation results of the baseline methods, and the results of our methods are shown in the third row.
Our methods generate better visual segmentation results than the baselines and are more similar to the ground-truth annotations, which are consistent with the quantitative results.

\section{Conclusion}\label{sect:conclusion}
This work aims to enable federated contrastive learning (FCL) for volumetric medical image segmentation with limited annotations.
Clients first learn a shared encoder on distributed unlabeled data and then a model is fine-tuned on annotated data.
Feature exchange is proposed to improve data diversity for contrastive learning while avoiding sharing raw data.
Global structural matching is developed to learn an encoder with unified representations among clients.
To reduce the communication cost of FCL, an optimized method FCLOpt that does not rely on negative samples is proposed. 
Based on FCLOpt, predictive target network update (PTNU) is developed by predicting the target network by fast forward to further reduce the communications of model downloading.
Distance prediction (DP) is proposed to remove the uploading of the target network.
The experimental results show significantly improved segmentation performance and labeling-efficiency compared with state-of-the-art techniques.

\section{Description of the Extensions}\label{sect:extension}
The original conference version of this paper was published on MICCAI 2021 proceedings \cite{wu2021federated}.
Compared with the original version, we made the following extensions in this manuscript.

\begin{enumerate}
    \item We added Section \ref{sect:FCL_reduce_comm} to describe the communication-optimized method FCLOpt. The FCL method introduced in the original MICCAI paper (described in Section \ref{sect:FCL_feature_share} in this manuscript) requires additional communication for feature sharing, aiming at improving the segmentation performance. We extend the original method by proposing an optimized method FCLOpt (Section \ref{sect:FCLOpt_overview} and Section \ref{sect:FCLOpt_training}) that does not rely on negative samples to eliminate the communication costs of feature sharing. 
    \item Based on FCLOpt, to further reduce the communications of model download, we propose the predictive target network update (PTNU) that predicts the target network by fast forward (Section \ref{sect:PTNU}).
    \item Based on PTNU, we propose the distance prediction (DP) to remove the uploading of the target network (Section \ref{sect:DP}).

    \item We added experiments for the extended methods. More specifically, we added experiments for the FCLOpt, FCL-PTNU, and FCL-PTNU-DP in Section \ref{sect:exp}. 
    The results of segmentation performance by local fine-tuning, federated fine-tuning, centralized fine-tuning, and transfer learning are added to Section \ref{sect:exp_local_finetune}, Section \ref{sect:exp_federated_finetune}, Section \ref{sect:exp_centralized_finetune}, and Section \ref{sect:exp_transfer_learning}, respectively.
    The results of communication cost are added to Section \ref{sect:exp_reduced_comm} to show the reduced communication by the added methods compared with the FCL method in the original paper.
    The visualization of segmentation results is added to Section \ref{sect:exp_visualization}.
    These results show the extended FCLOpt including the PTNU and DP methods effectively reduces the communication cost while preserving the segmentation performance of the FCL method in the original paper.
    \item In addition to the extended methods and corresponding experiments, in the experimental results of Section \ref{sect:exp}, we added two baseline methods FedMoCo and FedBYOL to the experimental results for comparison. We also added the evaluation protocol centralized fine-tuning with limited annotations, which is a standard evaluation protocol for generic self-supervised learning.
\end{enumerate}

\section*{Acknowledgments}
This work was supported in part by NSF CNS-2122320, CNS-212220, CNS-1822099, and CNS-2007302.

\bibliographystyle{IEEEtran}
\bibliography{IEEEexample}

\end{document}